\documentclass[%
 reprint,
 amsmath,amssymb,
 aps,
 prb,
]{revtex4-2}

\usepackage{graphicx}
\graphicspath{{figs/}}
\usepackage{dcolumn}
\usepackage{bm}
\usepackage[dvipsnames]{xcolor}
\usepackage{braket}
\usepackage{multirow}

\newcommand{\expo}[1]{{\rm e}^{#1}}
\newcommand{\iu}{{i\mkern1mu}}

\begin{document}

\title{Quasinormal modes of Floquet media slabs}
\author{Benjamin Vial}
\email{b.vial@imperial.ac.uk}
\affiliation{Department of Mathematics, Imperial College London, London SW7 2AZ, UK}
\author{Richard V. Craster}
\affiliation{Department of Mathematics,
    UMI 2004 Abraham de Moivre-CNRS,
    Department of Mechanical Engineering, Imperial College London, London SW7 2AZ, UK.}
\email{r.craster@imperial.ac.uk}
\date{\today}

\begin{abstract}
Exploiting non-Hermitian wave-matter interactions in time-modulated media to enable the dynamic control of electromagnetic waves requires advanced theoretical tools. In this article we bridge concepts from photonic quasinormal modes (QNMs) and time-varying metamaterials providing the foundation for designing dynamic optical devices with prescribed scattering properties. Establishing the QNM framework for slabs with time-periodic permittivity, and solving the associated nonlinear eigenvalue problem, allows us to derive the QNM expansion capturing the resonant features of the system. This reduced-order model enables highly efficient computation of scattered fields while revealing insight into how modulation couples to resonant modes, creating tailored gain-loss engineering. Our approach is validated through numerical experiments on time-modulated systems, and we design strategies to engineer tailored excitations selectively amplifying or suppressing specific modal contributions. 
\end{abstract}


\maketitle

\section{\label{sec:intro}Introduction}

Temporal modulation of material properties is emerging as a powerful paradigm for controlling electromagnetic wave propagation, enabling phenomena that are impossible with static media \cite{Galiffi2022photonics, PachecoPena2022, monticone2025}: time-varying metamaterials offer unprecedented opportunities to break reciprocity, achieve broadband amplification, and realize exotic wave phenomena through parametric processes~\cite{Caloz2019temporal, Yu2021complete}. However, the theoretical framework for analyzing such systems remains challenging, particularly when seeking to understand the interplay between temporal modulation and spatial resonances in finite geometries.

A powerful approach in static media for electromagnetic scattering relies on the concept of quasinormal modes (QNMs)---the natural resonances of open optical structures that govern their response to external excitation~\cite{Lalanne2018light, Kristensen2020modes}. These complex-frequency eigenmodes provide a basis for expanding scattered fields, enabling efficient computation and deep physical insight into resonant phenomena~\cite{ching1998, vial2014, sauvan2022}. This QNM framework has proven invaluable for static photonic systems, from plasmonic nanoparticles to photonic crystal cavities, where it helps understand the connection between geometry, material properties, and optical response~\cite{Bai2013efficient, Yan2018rigorous}.

The extension of QNM theory to time-modulated systems presents both conceptual and technical challenges. Unlike their static counterparts, Floquet media---materials with time-periodic properties---exhibit coupling between different frequency components, leading to parametric amplification, frequency conversion, and complex gain-loss dynamics~\cite{Cassedy1967dispersion, Felsen1970time}. 

Recent advances in temporal metamaterials have demonstrated remarkable control over wave propagation through time modulation. Parametric amplification has been observed in time-modulated transmission lines and photonic crystals~\cite{ValdezGarcia2024parametric, Planat2020Josephson}, while temporal boundaries have been exploited to manipulate wave momentum and energy~\cite{ValdezGarcia2024parametric}. Time-varying gain media have shown the ability to suppress radiative losses~\cite{hooper2025}, and temporal modulation has been used to achieve non-reciprocal wave propagation~\cite{tessierbrothelande2023}. However, a unified theoretical framework that can efficiently analyze and design such systems---particularly in finite geometries where spatial and temporal effects interplay---has remained elusive.

The challenge becomes particularly acute for slab geometries, which represent a fundamental building block for many photonic devices. Unlike infinite periodic structures that can be analyzed using Floquet-Bloch theory, finite slabs support leaky modes that radiate into the surrounding medium, creating an open scattering problem \cite{Wu2025reflections}. The temporal modulation couples these leaky modes, and understanding and controlling this coupling is essential for designing time-modulated optical devices with prescribed scattering properties.

In this article, we establish a comprehensive QNM framework for analyzing electromagnetic scattering from slabs with time-periodic permittivity. Our approach transforms the time-domain scattering problem into a nonlinear eigenvalue problem whose solutions---the Floquet quasinormal modes---capture both the spatial structure and temporal dynamics of the system. We develop a systematic method for computing these modes and demonstrate how they provide a reduced-order model for the scattered electromagnetic field.

Our theoretical framework is validated through comprehensive numerical experiments on representative time-modulated systems. We demonstrate excellent agreement between the QNM predictions and full-wave simulations, confirming the accuracy and efficiency of the approach. Furthermore, we present two concrete design strategies: one for achieving selective modal amplification and another for suppressing specific resonances while preserving others.


The paper is organized as follows. Section~\ref{sec:theory} describes the electromagnetic wave propagation for periodically time-modulated media. Section~\ref{sec:slab_theory} develops the theoretical framework for time-modulated slabs, including the scattering formulation, spectral problem and quasi-normal mode expansion based on the Keldysh theorem. In Section~\ref{sec:mode_control} we present mode-selective illumination strategies for controlling individual resonances, and finally Section~\ref{sec:conclusion} summarizes the main findings and discusses future directions. Numerical examples are used throughout the paper to illustrate our theoretical findings.

\section{Floquet Medium Theory}
\label{sec:theory}

We consider electromagnetic wave propagation in a medium with time-periodic permittivity $\varepsilon(t) = \varepsilon(t + T)$, where $T$ is the modulation period~\cite{zurita-sanchez2009}. For simplicity, we assume the permeability $\mu = \mu_0$ remains constant, though the formalism can be readily extended to include magnetic modulation. The time-varying nature of the medium fundamentally alters the wave propagation characteristics, enabling phenomena such as parametric amplification and non-reciprocal transmission that are absent in static media.

Maxwell's equations in such time-modulated media admit plane wave solutions of the form  
\begin{equation}
{E}({r}, t) = {E}(t) e^{i{k} \cdot {r}},
\end{equation}
where ${k}$ is the wave vector. Without loss of generality, we consider wave propagation in the $x$-direction, so that ${k} = k \hat{{x}}$ and the electric field becomes $E(x, t) = E(t) e^{ikx}$. 

Substituting this ansatz into Maxwell's equations and eliminating the magnetic field, we obtain the fundamental wave equation for time-modulated media:
\begin{equation}
\frac{d^{2}}{d t^{2}}[\varepsilon(t) E(t)] + k^{2} c^{2} E(t) = 0. \label{wave_1D_timemod}
\end{equation}

The solution to Eq.~(\ref{wave_1D_timemod}) is governed by the Floquet-Bloch theorem, which states that for time-periodic systems, the electric field can be written as:
\begin{equation}
E(t) = \bar{E}(\omega, t) e^{-i\omega t}, \label{floquet_form}
\end{equation}
where $\bar{E}(\omega, t)$ is $T$-periodic in time and $\omega$ is the angular frequency. 
To solve for the dispersion relation $\omega(k)$ and the corresponding electromagnetic modes, we expand both the permittivity and the periodic part of the electric field as Fourier series:
\begin{equation}
\varepsilon(t) = \sum_{q=-\infty}^{\infty} \varepsilon_{q} e^{iq\Omega t}\quad \text{and}
\quad \bar{E}(\omega, t) = \sum_{q=-\infty}^{\infty} e_{q}(\omega) e^{iq\Omega t}, \label{fourier_epsE}
\end{equation}
where $\Omega = 2\pi/T$ is the fundamental modulation frequency, and the Fourier coefficients $\varepsilon_q$ characterize the strength of the $q$-th harmonic in the permittivity modulation.

Substituting the Fourier expansions (\ref{fourier_epsE}) into the wave equation (\ref{wave_1D_timemod}) and equating coefficients of $\exp({ip\Omega t})$, we obtain an infinite system of coupled equations. This system can be written compactly as the matrix eigenvalue problem:
\begin{equation}
{N}(\omega) {e}(\omega) = k^{2} c^{2} {e}(\omega), \label{eq:mat_eig}
\end{equation}
where ${e}(\omega) = [e_{-M}, e_{-M+1}, \ldots, e_{M-1}, e_M]^T$ is the vector of Fourier coefficients (truncated to $2M+1$ terms for numerical implementation), and the matrix elements are given by:
\begin{equation}
N_{pq}(\omega) = (\omega - q\Omega)^{2} \varepsilon_{q-p}  \label{matrix_elements}
\end{equation}
for $p,q \in \{-M, -M+1, \ldots, M\}$.  
This eigenvalue problem determines the permissible wave vectors $k_n(\omega)$ for a frequency $\omega$, thereby defining the dispersion relation.

While the infinite medium analysis provides the foundation for understanding Floquet wave propagation, realistic photonic structures are finite and support leaky resonances that radiate into the surrounding medium. These resonances—the quasinormal modes—cannot be captured by the periodic boundary conditions implicit in the Floquet-Bloch analysis.

In the following sections, we extend this formalism to finite slab geometries, where the time-modulated region is bounded in space. This extension requires careful treatment of the boundary conditions and leads to a nonlinear eigenvalue problem whose solutions are the Floquet quasinormal modes that form the central focus of this work.

\section{Time-Modulated Slab}
\label{sec:slab_theory}
\subsection{Scattering Problem}
\label{subsec:scattering}

\begin{figure}[htbp]
    \centering
    \includegraphics[width=1\columnwidth]{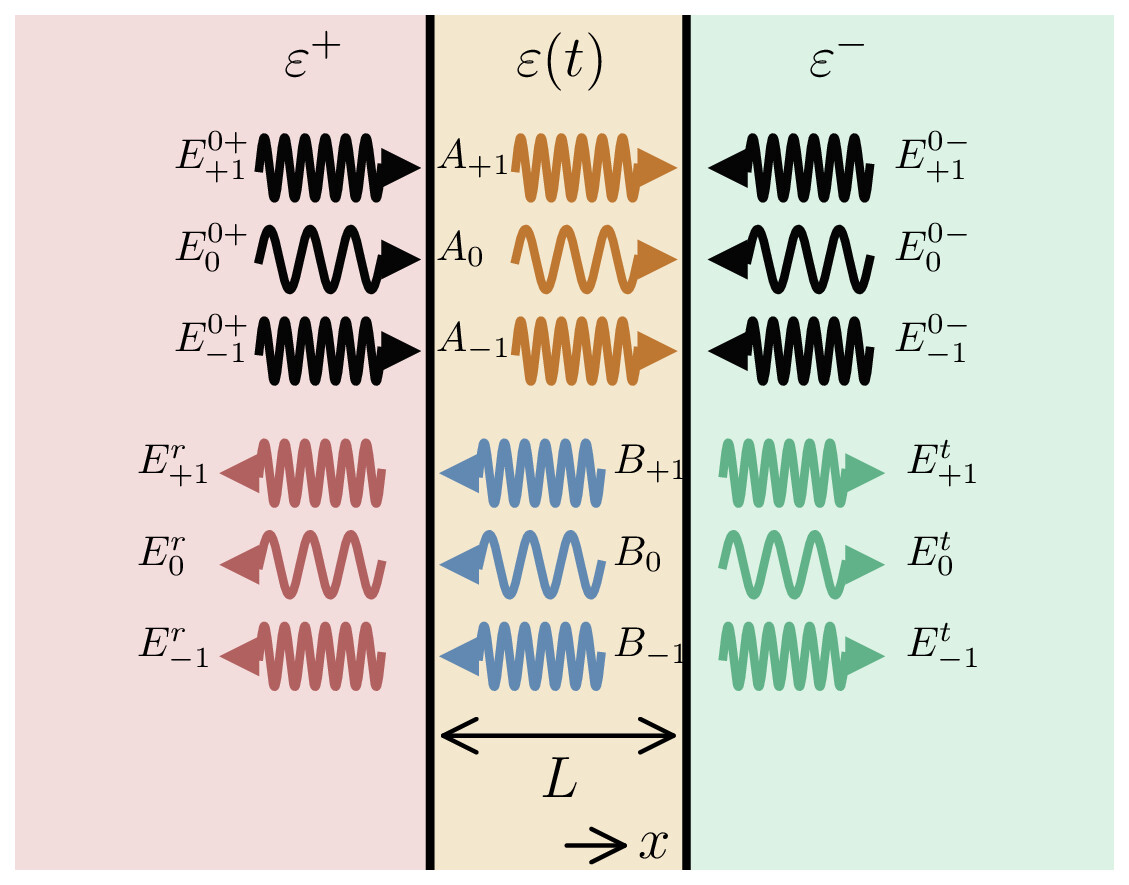}
    \caption{Schematic of a slab of thickness $L$ with time-periodic permittivity $\varepsilon(t)$ illuminated by electromagnetic waves from both sides. The slab is bounded by regions with permittivities $\varepsilon_-$ (left) and $\varepsilon_+$ (right). Multiple Floquet harmonics are shown: incident waves $E_q^{0\pm}$ from left and right, reflected waves $E_q^{r}$, and transmitted waves $E_q^{t}$, where $q = -1, 0, +1$ represents the harmonic index. The forward and backward propagating amplitudes within the slab are denoted as $A^q$ and $B^q$, respectively.}
    \label{fig:time_modulated_slab}
\end{figure}

We consider a slab of thickness $L$ occupying the region $0 < x < L$, with time-periodic dielectric function $\varepsilon(t) = \varepsilon(t+T)$ where $T = 2\pi/\Omega$ is the modulation period. The surrounding media have constant permittivities $\varepsilon^{-}$ (for $x < 0$) and $\varepsilon^{+}$ (for $x > L$), as illustrated in Fig.~(\ref{fig:time_modulated_slab}). This configuration creates a spatio-temporal interface problem where the finite extent of the modulated region gives rise to leaky resonances—the Floquet quasinormal modes that form the focus of our analysis.

For normal incidence, we assume a superposition of plane waves with fundamental frequency $\omega$ striking the interfaces from both sides. Due to the time-periodic nature of the slab, the incident, reflected, and transmitted fields all contain multiple frequency harmonics $\omega - q\Omega$ where $q \in \mathbb{Z}$.

The incident electromagnetic fields are expressed as:
\begin{equation}
    E_{\mathrm{inc}}(x, t) = \sum_{q=-\infty}^{\infty}\left[E^{0+}_{q} e^{ik^{+} x} 
    + E^{0-}_{q} e^{-ik^{-} x}\right]e^{-i(\omega - q \Omega) t}, \label{eq:incident_E}
\end{equation}
\begin{equation}
    H_{\mathrm{inc}}(x, t) = \sum_{q=-\infty}^{\infty}\left[H^{0+}_{q} e^{ik^{+} x} 
    + H^{0-}_{q} e^{-ik^{-} x}\right]e^{-i(\omega - q \Omega) t}, \label{eq:incident_H}
\end{equation}
where $k^{\pm}= \sqrt{\varepsilon^{\pm}} \omega / c$ are the wave vectors in the surrounding media. The superscripts $+$ and $-$ denote waves incident from the left and right sides, respectively.

Inside the time-modulated slab, the electromagnetic fields are superpositions of the Floquet normal modes derived in Section~\ref{sec:theory}:
\begin{equation}
    \begin{split}
    E_{\mathrm{sl}}(x, t) = &\sum_{p=1}^{\infty} \sum_{q=-\infty}^{\infty}\left[A_{p} e^{ik_{p}(\omega) x} \right. \\
    &\left. + B_{p} e^{-ik_{p}(\omega) x}\right] e_{pq}(\omega) e^{-i(\omega - q \Omega) t}, \label{eq:slab_E}
    \end{split}
\end{equation}
\begin{equation}
    \begin{split}
    H_{\mathrm{sl}}(x, t) = &\sum_{p=1}^{\infty} \sum_{q=-\infty}^{\infty}\left[C_{pq} e^{ik_{p}(\omega) x} \right. \\
    &\left. + D_{pq} e^{-ik_{p}(\omega) x}\right] e_{pq}(\omega) e^{-i(\omega - q \Omega) t}, \label{eq:slab_H}
    \end{split}
\end{equation}
where $k_p(\omega)$ and $e_{pq}(\omega)$ are the eigenvalues and eigenfunctions of the matrix problem (\ref{eq:mat_eig}), and $A_p$, $B_p$ are the modal amplitudes to be determined from boundary conditions.

The reflected and transmitted fields in the exterior regions are:
\begin{gather}
    E_{r}(x, t) = \sum_{q} E_{q}^{r} e^{-i\left[k_{q}^{r}(\omega) x + (\omega - \Omega q) t\right]}, \label{eq:reflected_E} \\
    H_{r}(x, t) = -\sum_{q} H_{q}^{r} e^{-i\left[k_{q}^{r}(\omega) x + (\omega - \Omega q) t\right]}, \label{eq:reflected_H}
\end{gather}
\begin{align}
    & E_{t}(x, t) = \sum_{q} E_{q}^{t} e^{i\left[k_{q}^{t}(\omega)(x - L) - (\omega - \Omega q) t\right]}, \label{eq:transmitted_E} \\
    & H_{t}(x, t) = \sum_{q} H_{q}^{t} e^{i\left[k_{q}^{t}(\omega)(x - L) - (\omega - \Omega q) t\right]}, \label{eq:transmitted_H}
\end{align}
where $k_{q}^{r,t}(\omega) = \sqrt{\varepsilon^{+,-}}(\omega - \Omega q) / c$ account for the frequency shifts of the harmonics.

From Maxwell's equations, the electric and magnetic field amplitudes are related by:
\begin{gather}
    H^{0\pm}_{q} = \sqrt{\frac{\varepsilon_{0} \varepsilon^{\pm}}{\mu_{0}}} E^{0\pm}_{q}, \quad H_{q}^{r, t} = \sqrt{\frac{\varepsilon_{0} \varepsilon^{+,-}}{\mu_{0}}} E_{q}^{r, t}, \label{eq:maxwell_relations} \\
    \begin{bmatrix} C_{pq} \\ D_{pq} \end{bmatrix} = \sqrt{\frac{\varepsilon_{0}}{\mu_{0}}} \frac{k_{p}(\omega) c}{\omega - \Omega q} \begin{bmatrix} A_{p} \\ -B_{p} \end{bmatrix}. \label{eq:amplitudes_relation}
\end{gather}

The continuity of tangential electric and magnetic fields at the interfaces $x = 0$ and $x = L$ provides the boundary conditions:
\begin{align}
    & E_{\mathrm{inc}}(0, t) + E_{r}(0, t) = E_{\mathrm{sl}}(0, t), \label{eq:boundary_E_0} \\
    & H_{\mathrm{inc}}(0, t) + H_{r}(0, t) = H_{\mathrm{sl}}(0, t), \label{eq:boundary_H_0} \\
    & E_{\mathrm{sl}}(L, t) = E_{t}(L, t), \label{eq:boundary_E_L}\\
    & H_{\mathrm{sl}}(L, t) = H_{t}(L, t). \label{eq:boundary_H_L}
\end{align}

Applying the boundary conditions and eliminating the internal modal amplitudes $A_p$ and $B_p$, we obtain a linear system relating the incident and scattered field amplitudes:
\begin{gather}
E^{0+}_{q} + E_{q}^{r} = \sum_{p} e_{pq}(\omega) \left[A_{p} + B_{p}\right], \label{eq:linear1}\\
(E^{0+}_{q} - E_{q}^{r} )\sqrt{\varepsilon^{+}} = \sum_{p} \frac{e_{pq}(\omega) k_{p}(\omega) c}{\omega - q \Omega} \left[A_{p} - B_{p}\right], \label{eq:linear2}\\
E^{0-}_{q} + E_{q}^{t} = \sum_{p} e_{pq}(\omega) \left[A_{p} e^{ik_{p}(\omega) L} + B_{p} e^{-ik_{p}(\omega) L}\right], \label{eq:linear3}\\
\begin{split}
(E_{q}^{t} - E^{0-}_{q})\sqrt{\varepsilon^{-}} = &\sum_{p} \frac{e_{pq}(\omega) k_{p}(\omega) c}{\omega - q \Omega} \\
&\left[A_{p} e^{ik_{p}(\omega) L} - B_{p} e^{-ik_{p}(\omega) L}\right]. \label{eq:linear4}
\end{split}
\end{gather}
This system can be written compactly in matrix form as:
\begin{equation}
S(\omega)\ket{ \phi} = \ket{\phi^0}
\label{eq:slab_scatt}
\end{equation}
with the scattering matrix $S = M W^{-1}$ and the block matrices $M$ and $W$ are defined as
\begin{equation}
M_{pq} = \frac{1}{2}
    \begin{pmatrix}
    1 + \hat{n}_{p}/n_q^+   &   1 - \hat{n}_{p}/n_q^+  \\
    \left( 1 - \hat{n}_{p}/n_q^- \right) \expo{\iu k_pL} & \left(   1 + \hat{n}_{p}/n_q^- \right) \expo{-\iu k_pL}
    \end{pmatrix}e_{p q}
    \label{eq:slab_matrix1}
\end{equation}
\begin{equation}
W_{pq} = 
    \begin{pmatrix}
    1   &   1  \\
     \expo{\iu k_pL} & \expo{-\iu k_pL}
    \end{pmatrix}e_{p q}
\end{equation}
where $\hat{n}_p = k_p c /\omega$, $n_q^{\pm} = (1-q\Omega/\omega) n^\pm$ and $n^\pm = \sqrt{\varepsilon^{\pm}}$. 
This relates the input and output vectors $\ket{\phi^0}$ and $\ket{\phi}$ defined as 
\begin{equation*}
\ket{\phi^0} = 
    \begin{pmatrix}
    E^{0+}_{q} \\[0.7ex]
    E^{0-}_{q}
    \end{pmatrix}, \qquad 
\ket{\phi} = 
    \begin{pmatrix}
    E_{q}^r \\[0.7ex]
    E_{q}^t
    \end{pmatrix},
\end{equation*}

The scattering matrix ${S}(\omega)$ completely characterizes the linear response of the time-modulated slab. Its frequency dependence encodes both the temporal modulation effects and the spatial resonances of the finite geometry.

\subsection{Spectral Problem}
\label{subsec:qnm_eigenvalue}

The quasinormal modes of the time-modulated slab correspond to the complex frequencies $\omega_n$ at which the homogeneous version of the scattering problem has nontrivial solutions. These occur when the scattering matrix becomes singular:
\begin{align}
{S}(\omega_p)|\phi_p\rangle &= 0, \label{eq:slab_evp} \\
\langle\psi_p| {S}(\omega_p) &= 0, \label{eq:slab_evp_left}
\end{align}
where $|\phi_p\rangle$ and $\langle\psi_p|$ are the right and left eigenvectors, respectively. 
We solve this nonlinear eigenvalue problem using Newton's method combined with generalized Rayleigh quotient iteration~\cite{ruheAlgorithmsNonlinearEigenvalue1973,guttelNonlinearEigenvalueProblem2017}. This iterative approach efficiently handles the frequency dependence of the scattering matrix while maintaining numerical stability.

Furthermore, the eigenvectors are normalized using the generalized biorthogonality condition:
\begin{equation}
    \langle \psi_p | \phi_q \rangle = \delta_{pq} := \begin{cases}
        \langle \psi_p | \frac{{S}(\omega_p)-{S}(\omega_q)}{\omega_p - \omega_q} | \phi_q \rangle, & \text{if } \omega_p \neq \omega_q \\[1ex]
       \langle \psi_p |\frac{\partial {S}}{\partial \omega}(\omega_p) | \phi_p \rangle, & \text{if } \omega_p = \omega_q.
    \end{cases}
    \label{eq:ortho}
\end{equation}

To validate our theoretical framework and demonstrate the effects of temporal modulation on the quasi-normal mode spectrum, we present numerical calculations for a representative Floquet medium slab. We consider a slab of thickness $L$ in vacuum subject to sinusoidal temporal modulation $\varepsilon(t) = \varepsilon_0 + \Delta \varepsilon \sin(\Omega t)$, with $\varepsilon_0 = 50$, $\Delta\varepsilon = 20$, $L = 1.5\mu m$ and $\Omega=c/L$ corresponding to a frequency of $31.8$THz. In the following numerical examples, we restrict the truncation of harmonics up to $q=\pm3$.

\begin{figure}[htbp]
\centering
\includegraphics[width=1\columnwidth]{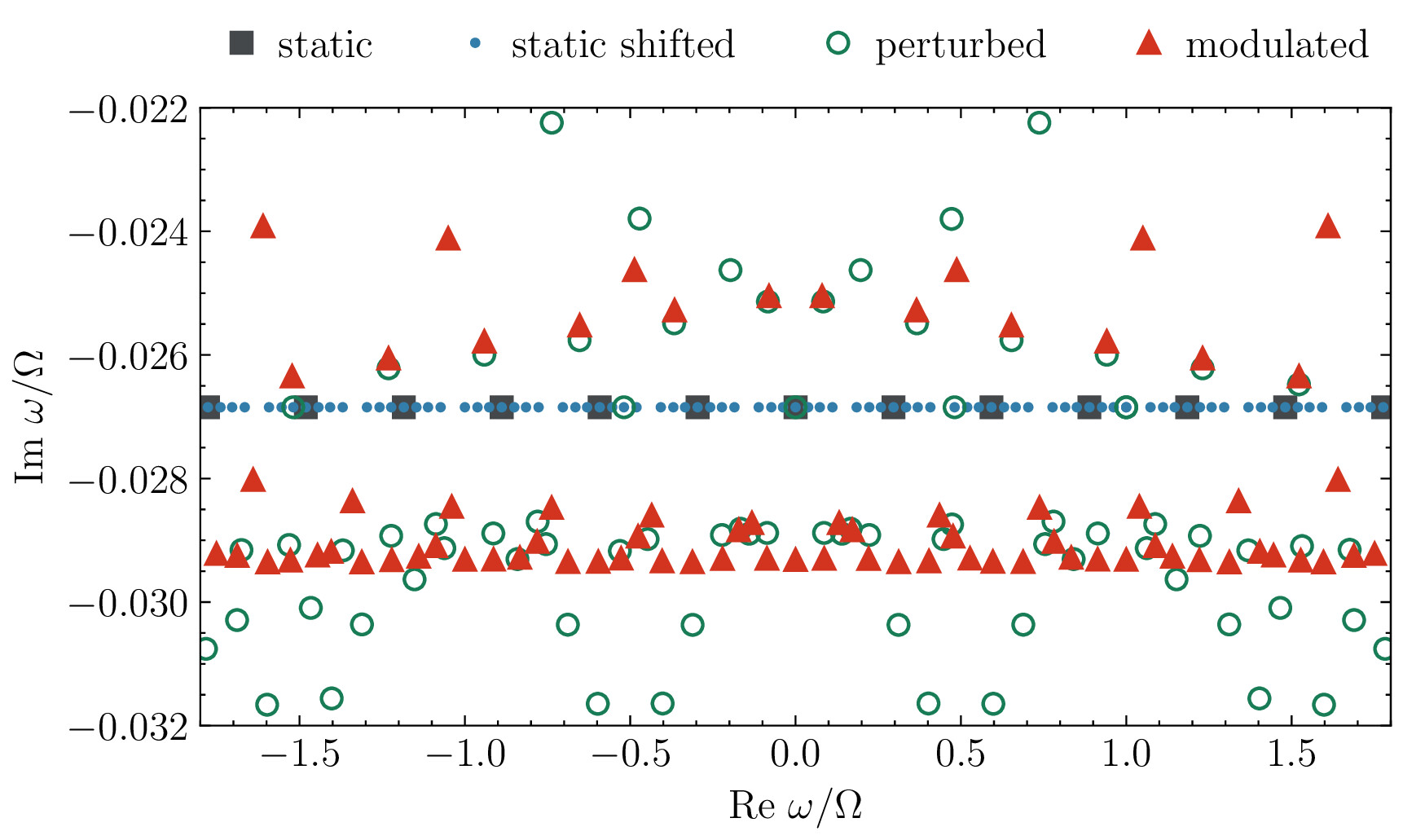}
\caption{Complex eigenfrequency spectrum of the Floquet medium slab showing the evolution from static to fully modulated regimes. Black squares: static unmodulated modes; blue dots: static modes shifted by integer multiples of $\Omega$; green circles: first-order perturbation theory results; red triangles: full numerical solution with temporal modulation. The spectrum demonstrates how temporal modulation couples modes across different frequency sidebands.}
\label{fig:spectrum}
\end{figure}

Figure~(\ref{fig:spectrum}) displays the complex eigenfrequency spectrum in the normalized frequency plane. The spectrum reveals four distinct regimes that illustrate the progressive effects of temporal modulation. The static modes (black squares, see Appendix~\ref{sec:static}) represent the unmodulated slab eigenvalues $\widetilde{\omega}_q$, which form the baseline reference spectrum. When temporal modulation is introduced, the empty temporal lattice approximation (see Appendix~\ref{sec:empty}) produces static shifted modes (blue dots) given by $\widetilde{\omega}_{p,q} = \widetilde{\omega}_q + p\Omega$, representing integer shifts of the static modes by multiples of the modulation frequency. First-order perturbation theory (green circles, see Appendix~\ref{sec:pert}) shows small deviations from the shifted static modes due to weak coupling effects, while the full numerical solution (red triangles) captures all temporal coupling effects and demonstrates the complete impact of Floquet modulation. 
The transition from perturbative results to the fully modulated spectrum reveals that strong coupling effects become significant, requiring the complete Floquet treatment beyond simple perturbation theory. This spectral analysis confirms that temporal modulation creates new pathways for energy coupling.

\begin{table*}[htbp]
\centering
\caption{Comparison of complex eigenfrequencies obtained from solving Eq.~(\ref{eq:slab_evp}) (NLEVP) and finite element method (FEM). All frequencies are normalized with respect to $\Omega$. Relative errors are computed separately for real and imaginary parts as $|\text{Re}(\omega_{\text{NLEVP}}) - \text{Re}(\omega_{\text{FEM}})|/|\text{Re}(\omega_{\text{FEM}})| \times 100\%$ and similarly for imaginary parts.}
\label{tab:fem_comparison}
\begin{tabular}{@{}c|cc|cc|cc@{}}
\hline
\hline
\multirow{2}{*}{Mode} & \multicolumn{2}{c|}{NLEVP} & \multicolumn{2}{c|}{FEM} & \multicolumn{2}{c}{Relative Error (\%)} \\
& Re & Im & Re & Im & Re & Im \\
\hline
1  & 0.132009 & $-0.028722$ & 0.132009 & $-0.028722$ & $3.48e\text{-}05$ & $3.13e\text{-}05$ \\
2  & 0.171810 & $-0.028836$ & 0.171810 & $-0.028836$ & $5.35e\text{-}06$ & $1.52e\text{-}04$ \\
3  & 0.204133 & $-0.024832$ & 0.204133 & $-0.024832$ & $1.87e\text{-}04$ & $4.79e\text{-}05$ \\
4  & 0.220921 & $-0.029302$ & 0.220922 & $-0.029303$ & $6.07e\text{-}04$ & $3.83e\text{-}03$ \\
5  & 0.311828 & $-0.029355$ & 0.313238 & $-0.030307$ & $4.50e\text{-}01$ & $3.14e\text{+}00$ \\
6  & 0.365878 & $-0.025279$ & 0.365878 & $-0.025279$ & $4.19e\text{-}05$ & $2.15e\text{-}06$ \\
7  & 0.403375 & $-0.029340$ & 0.402187 & $-0.031571$ & $2.95e\text{-}01$ & $7.07e\text{+}00$ \\
8  & 0.435185 & $-0.028599$ & 0.435185 & $-0.028599$ & $1.10e\text{-}05$ & $1.67e\text{-}04$ \\
9  & 0.476284 & $-0.028939$ & 0.476285 & $-0.028938$ & $9.20e\text{-}05$ & $6.32e\text{-}03$ \\
10 & 0.487355 & $-0.024626$ & 0.487356 & $-0.024626$ & $1.21e\text{-}04$ & $6.75e\text{-}05$ \\
11 & 0.527468 & $-0.029296$ & 0.527467 & $-0.029294$ & $8.00e\text{-}05$ & $6.38e\text{-}03$ \\
12 & 0.596660 & $-0.029349$ & 0.597848 & $-0.031580$ & $1.99e\text{-}01$ & $7.06e\text{+}00$ \\
13 & 0.652814 & $-0.025524$ & 0.652814 & $-0.025524$ & $1.29e\text{-}05$ & $1.17e\text{-}05$ \\
14 & 0.688185 & $-0.029350$ & 0.686775 & $-0.030302$ & $2.05e\text{-}01$ & $3.14e\text{+}00$ \\
15 & 0.737713 & $-0.028475$ & 0.737713 & $-0.028475$ & $1.93e\text{-}05$ & $1.70e\text{-}05$ \\
16 & 0.769460 & $-0.024423$ & 0.769461 & $-0.024423$ & $1.14e\text{-}04$ & $9.65e\text{-}05$ \\
17 & 0.781500 & $-0.029027$ & 0.781499 & $-0.029028$ & $1.69e\text{-}04$ & $3.90e\text{-}03$ \\
\hline
\hline
\end{tabular}
\end{table*}

To validate the accuracy of our QNM framework, we compare the computed eigenfrequencies with those obtained from a finite element method (FEM) implementation detailed in Appendix~\ref{app:fem}. 
Table~\ref{tab:fem_comparison} presents a detailed comparison of the complex eigenfrequencies for the first 17 quasinormal modes of a time-modulated slab, showing both real and imaginary parts separately along with their respective relative errors. The 
results from the resolution of  the nonlinear eigenvalue problem~(\ref{eq:slab_evp}) (NLEVP) demonstrates excellent agreement with the FEM results for the majority of modes. Most modes exhibit exceptionally small relative errors, typically below $10^{-3}\%$ for both real and imaginary parts, indicating near-perfect agreement between the two methods. The modes with larger discrepancies highlight the numerical challenges associated with strongly coupled Floquet systems, particularly dealing with radiation conditions in FEM modal problems.

\subsection{Quasi-Normal Mode Expansion}
We employ the Keldysh theorem \cite{keldysh_completeness_1971}, a foundational result in the theory of non-self-adjoint operators that provides a completeness relation for a broad class of nonlinear eigenvalue problems. This theorem is particularly powerful for analyzing differential operators commonly encountered in physics and engineering, where non-self-adjoint and dispersive systems frequently arise. It has been successfully applied in optical systems \cite{zollaPhotonicsHighlyDispersive2018, truongContinuousFamilyExact2020} and elasticity \cite{vial2024a}. 

Using the Keldysh theorem, we can express the inverse of the scattering matrix as
\begin{equation}
S^{-1}(\omega) = \sum_{p} \frac{\ket{\phi_p} \bra{\psi_p}}{\omega-\omega_p} + R(\omega),
\label{eq:keldysh1}
\end{equation}
where $R(\omega)$ is a holomorphic function representing the non-resonant background contribution. We approximate this background term by a polynomial expansion:
\begin{equation}
R(\omega) \simeq \sum_{q=0}^{Q} R_q\omega^q.
\label{eq:pol}
\end{equation}
Consequently, the field solution can be approximated as
\begin{equation}
\ket{\phi} = S^{-1}\ket{\phi^0} \simeq \sum_{p=0}^{P} \alpha_p \ket{\phi_p} + \sum_{q=0}^{Q} \beta_q\omega^q,
\label{eq:qnmexp}
\end{equation}
where the coupling coefficient $\alpha_p = \frac{\braket{\psi_p | \phi^0}}{\omega-\omega_p}$ quantifies the strength of resonant excitation of the $p$-th mode by the incident field. The unknown coefficients $\beta_q = R_q\ket{\phi^0}$ are determined by evaluating Eq.~\eqref{eq:slab_scatt} at $Q+1$ distinct frequencies and solving the resulting linear system derived from Eq.~\eqref{eq:qnmexp}.

\begin{figure}[htbp]
\centering
\includegraphics[width=\columnwidth]{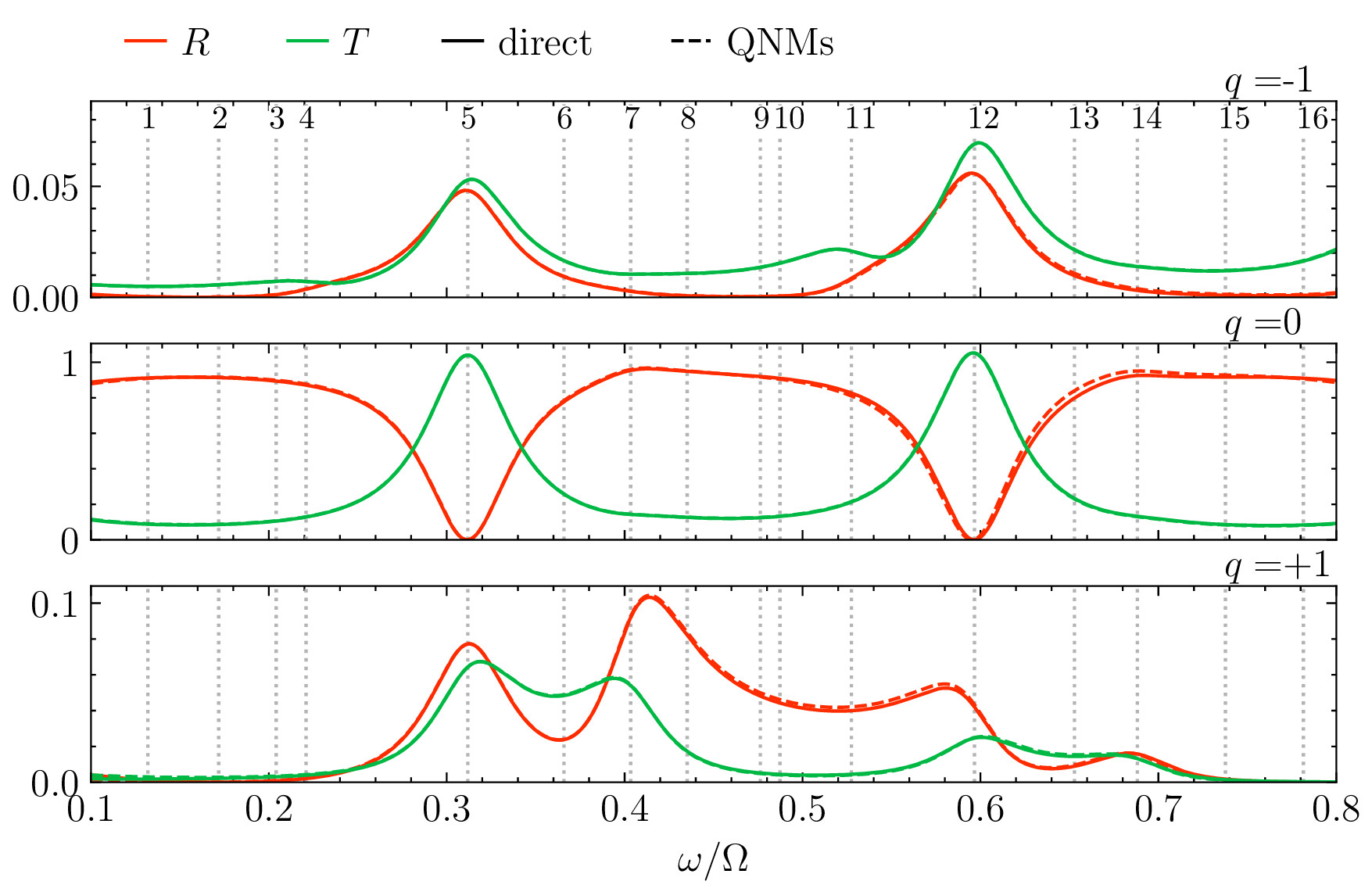}
\caption{Validation of the quasi-normal mode expansion against direct numerical calculations. Reflection ($R$) and transmission ($T$) coefficients are shown for Floquet harmonics $q = -1, 0, 1$ as functions of normalized frequency $\omega/\Omega$. Solid lines: direct numerical solution; dashed lines: quasi-normal mode expansion results. The excellent agreement demonstrates the accuracy and efficiency of the QNM approach.}
\label{fig:comparison}
\end{figure}

To validate the accuracy of our quasi-normal mode expansion approach, we compare the scattering coefficients obtained from the QNM method with direct numerical calculations. Figure~(\ref{fig:comparison}) shows the reflection ($R$) and transmission ($T$) coefficients as functions of normalized frequency $\omega/\Omega$ for three different Floquet harmonics: $q = -1, 0, +1$. The solid lines represent direct numerical solutions of the full scattering problem, while the dashed lines show the results obtained using the quasi-normal mode expansion from Eq.~\eqref{eq:qnmexp} with a constant approximation for the residual term $R$.

The agreement between the two methods is excellent across the chosen frequency range and harmonics, demonstrating the validity and accuracy of the QNM approach. For the fundamental harmonic $q = 0$, both reflection and transmission exhibit rich spectral features with multiple resonances and anti-resonances that are perfectly captured by the QNM expansion. The $q = -1$ and $q = +1$ harmonics show different spectral behaviors, with the $q = +1$ case displaying particularly strong resonant features around $\omega/\Omega \approx 0.2$ and $\omega/\Omega \approx 0.6$. These frequency-dependent scattering properties directly reflect the underlying quasi-normal mode structure and illustrate how temporal modulation enables selective frequency conversion between different harmonics.

The precision of the QNM reconstruction validates our theoretical framework and confirms that the quasi-normal mode expansion provides an efficient and accurate method for analyzing the electromagnetic response of Floquet media slabs. Once the spectral problem is solved, this approach significantly reduces computational complexity compared to direct frequency-domain calculations as one will not have to solve Eq.~(\ref{eq:slab_scatt}) when changing the incoming field parameters, while maintaining high accuracy across the selected spectral range.

\section{Mode-Selective Illumination}
\label{sec:mode_control}

The quasi-normal mode expansion framework enables sophisticated control over the electromagnetic response by designing incident fields that selectively excite or suppress specific eigenmodes. This capability opens new possibilities for tailoring the scattering properties and achieving targeted modal interactions in Floquet media slabs.

\subsection{Selective Excitation of a Single Mode}

A powerful strategy for controlling modal interactions involves designing source distributions that selectively couple to a single target eigenmode $p_0$ while suppressing all others. To achieve this, we define the $N \times (N-1)$ rectangular matrix $K_{p_0}$ whose columns comprise all left eigenvectors $\bra{\psi_q}$ except for the target mode $\bra{\psi_{p_0}}$. By constructing the incident field $\ket{\phi^0}$ to lie in the null space of $K_{p_0}$, we ensure that the coupling coefficients $\alpha_p \sim \braket{\psi_p | \phi^0} = 0$ for all $p \neq p_0$ by design.

This selective excitation scheme effectively isolates the contribution of the target mode, allowing for precise control over the electromagnetic response. The resulting field distribution is dominated by the single eigenmode $p_0$, providing a clean experimental signature of the mode's properties and enabling targeted manipulation of the scattering characteristics.

\subsection{Selective Suppression of a Single Mode}

Conversely, we can design incident fields that suppress the contribution of a specific unwanted mode $p_0$ while maintaining the excitation of all other modes. Using the expansion in Eq.~\eqref{eq:qnmexp} and exploiting the biorthogonality properties of the eigenmodes, we construct the incident field as
\begin{equation}
\ket{\phi^0} = \sum_{p \neq p_0} a_p \frac{S(\omega_{p_0}) - S(\omega_p)}{\omega_{p_0} - \omega_p} \ket{\phi_p},
\end{equation}
where the $a_p$ are arbitrary complex coefficients that determine the relative contributions of the remaining modes. By construction, this choice ensures that $\alpha_{p_0} \sim \braket{\psi_{p_0} | \phi^0} = 0$, effectively eliminating the contribution of mode $p_0$ from the total response.

This suppression technique is particularly valuable when certain modes exhibit undesirable properties, such as excessive losses or unwanted coupling to other degrees of freedom. By systematically removing these contributions, we can engineer the overall electromagnetic response to meet specific design requirements while maintaining the beneficial effects of the remaining modes.

The ability to selectively excite or suppress individual quasi-normal modes provides a powerful tool for electromagnetic design in temporally modulated media, enabling precise control over frequency conversion, scattering directionality, and energy transfer pathways.

\begin{figure}[htbp]
\centering
\includegraphics[width=1\columnwidth]{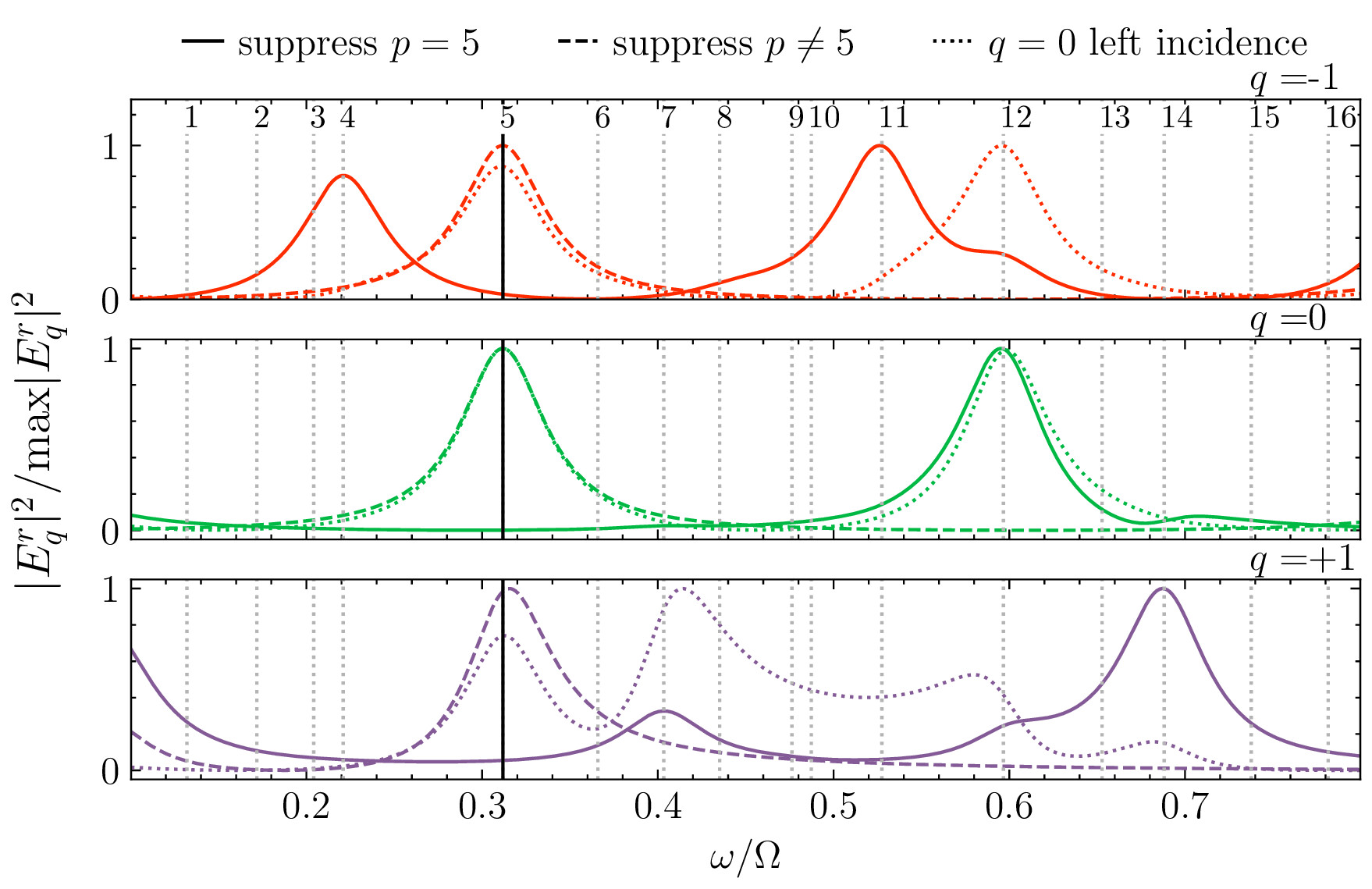}
\caption{Demonstration of mode-selective illumination control showing normalized reflected field intensity $|E_r^q|^2/\max|E_r^q|^2$ versus normalized frequency $\omega/\Omega$ for Floquet harmonics $q = -1, 0, +1$. Solid lines: suppression of mode $p = 5$; dashed lines: suppression of mode $p = 6$; dotted lines: conventional left incidence. The controlled illumination schemes selectively modify the spectral response, demonstrating effective mode control.}
\label{fig:comparison_control}
\end{figure}

Figure~\ref{fig:comparison_control} demonstrates the effectiveness of our mode control strategies through numerical simulations. The figure shows the normalized reflected field intensity $|E_r^q|^2/\max|E_r^q|^2$ as a function of normalized frequency $\omega/\Omega$ for the three Floquet harmonics $q = -1, 0, +1$. Three different illumination scenarios are compared: suppression of mode $p = 5$ (solid lines), suppression of modes $p\neq 5$ (dashed lines), and conventional left incidence in the $0^{\rm th}$ harmonic without mode control (dotted lines).

The results clearly illustrate the dramatic effect of selective mode suppression on the scattering response. For conventional left incidence, the reflected field exhibits multiple resonant peaks across all three harmonics, corresponding to the excitation of quasi-normal mode of the system (dotted lines). However, when specific modes are suppressed, characteristic spectral features are selectively eliminated or significantly modified.

The suppression of mode $p = 5$ (solid lines) produces notable changes in the spectral response, particularly evident in the $q = 0$ and $q = +1$ harmonics where certain resonant features are diminished compared to the uncontrolled case. Similarly, suppressing mode $p \neq 5$ (dashed lines) results in a different modification pattern, demonstrating that each mode contributes distinct spectral signatures that can be individually controlled. 
These observations are supported by the computed excitation coefficients $\alpha_p$, which show values for $p_0$ that are orders of magnitude larger or smaller than those of the other modes.

These results validate the theoretical framework for mode-selective illumination and demonstrate its practical utility for engineering the electromagnetic response of Floquet media. The ability to selectively enhance or suppress specific spectral features opens new possibilities for designing frequency-selective devices, achieving targeted scattering properties, and controlling energy flow in temporally modulated photonic systems.

\section{Parametric Amplification}
\label{sec:ampl}

\begin{figure}[htbp]
    \centering
    \includegraphics[width=1\columnwidth]{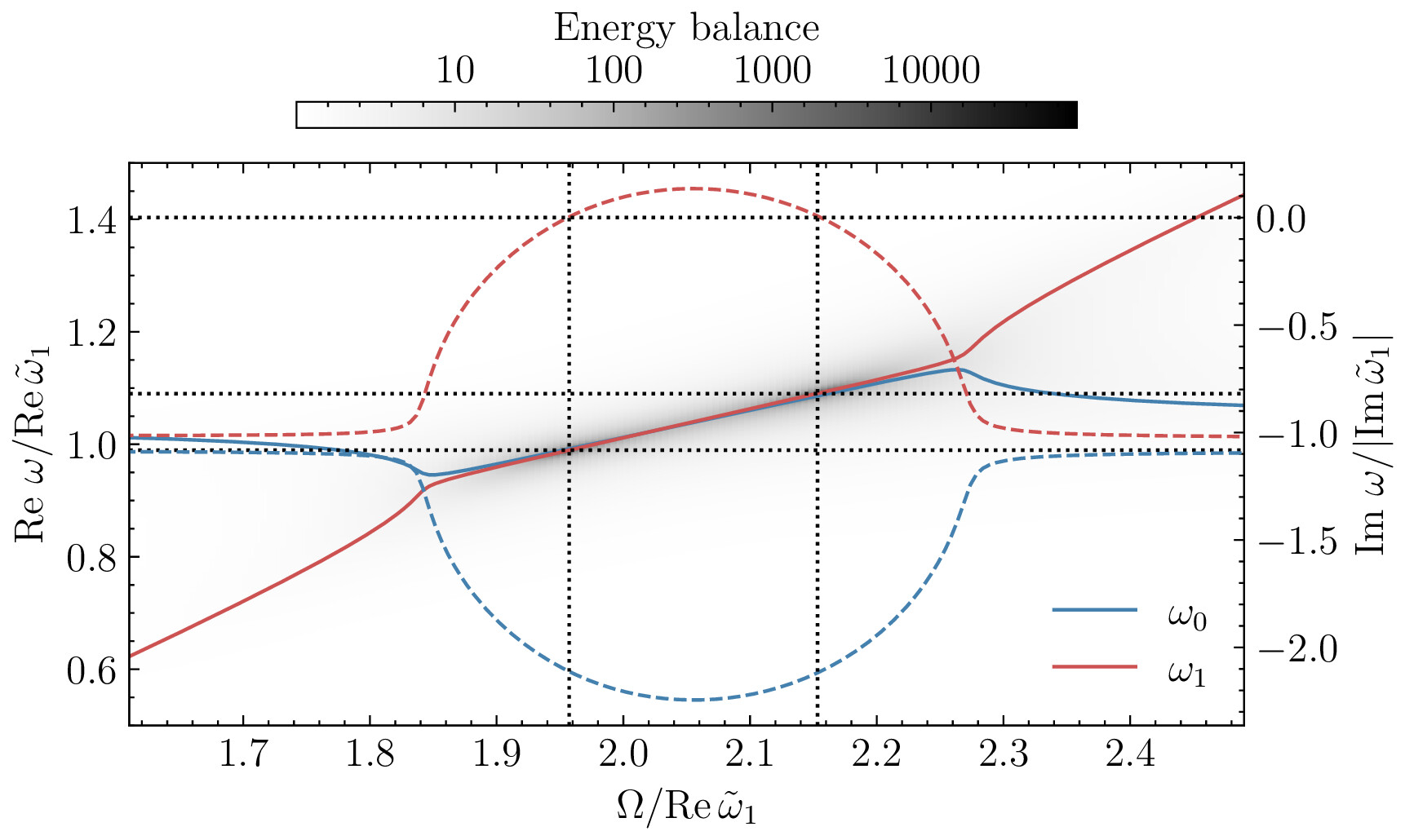}
    \caption{
        Evolution of quasinormal mode frequencies under time-periodic modulation. 
        The lines show normalized eigenfrequencies (real part: solid lines, imaginary part: dashed lines) as a function of modulation frequency $\Omega / \mathrm{Re}\,\tilde{\omega}_1$, where $\omega_0$ and $\omega_1$ denote resonant frequencies. The left and right vertical axes correspond to the real and imaginary parts respectively. 
        Color intensity represents energy amplification (log scale), revealing optimal modulation at $\Omega \approx 2.0\,\mathrm{Re}\,\tilde{\omega}_1$ where minimal leakage coincides with $10^4$ energy gain.
    }
    \label{fig:mode_evolution}
\end{figure}

The behavior of quasinormal modes under time-periodic modulation reveals rich physics associated with parametric coupling and the emergence of exceptional point dynamics \cite{Miri2019}. Figure~(\ref{fig:mode_evolution}) presents a comprehensive analysis of how two eigenfrequencies (labeled $\omega_0$ and $\omega_1$) evolve as a function of the normalized modulation frequency $\Omega/\mathrm{Re}\,\tilde{\omega}_1$, where $\tilde{\omega}_1$ represents the fundamental resonance of the unmodulated slab.

The real parts of both eigenfrequencies exhibit clear anticrossing behavior around $\Omega \approx 2.0\,\mathrm{Re}\,\tilde{\omega}_1$, indicating strong parametric coupling between the modes. This anticrossing is characteristic of level repulsion in non-Hermitian systems~\cite{Heiss2012}, where the modulation provides the coupling mechanism between otherwise orthogonal eigenstates. 

Simultaneously, the imaginary component of mode $\omega_1$ crosses zero near the optimal modulation frequency, transitioning from lossy to amplifying behavior, while the imaginary part of $\omega_0$ remains negative throughout the frequency range. This zero-crossing of $\operatorname{Im}\omega_1$ indicates that radiative losses are completely compensated for this particular mode precisely when parametric coupling is maximized.

The background colormap reveals that maximum energy amplification occurs precisely at the frequency where mode coupling is strongest and radiation losses are minimized. This correlation between coupling strength and energy gain demonstrates the critical role of proper frequency matching in achieving efficient parametric amplification in time-modulated photonic systems.

\section{Conclusion}
\label{sec:conclusion}

We have developed a comprehensive quasinormal mode (QNM) framework for analyzing electromagnetic scattering in slabs with time-periodic permittivity. By formulating the problem as a nonlinear eigenvalue problem, we obtained Floquet quasinormal modes that capture both the spatial and temporal structure of the system. This approach enables a reduced-order representation of the scattered field, revealing how temporal modulation selectively couples and modifies resonant modes.

Our theoretical formulation provides new insight into the interplay between temporal modulation and spatial resonances in open systems. In particular, we demonstrate how parametric amplification and mode hybridization emerge naturally from the QNM analysis. These effects can be harnessed to design targeted excitation strategies that amplify or suppress specific modal contributions.

The proposed framework was validated through numerical simulations, showing excellent agreement with full-wave solutions and confirming the accuracy and efficiency of the QNM-based approach. We also presented two design strategies for dynamic mode control, paving the way toward programmable photonic devices based on time modulation.

Beyond slab geometries, the tools and concepts introduced here lay the groundwork for a general theory of time-modulated photonic systems. They offer a unified perspective that connects non-Hermitian physics, temporal metamaterials, and modal analysis. We anticipate that this framework will find broad applications in the design of active optical components, non-reciprocal systems, and parametric devices operating far from equilibrium.

\begin{acknowledgments}
BV and RVC are supported by the H2020 FET-proactive Metamaterial Enabled Vibration Energy Harvesting (MetaVEH) project under Grant Agreement No. 952039. RVC is also funded by UK Research and Innovation (UKRI) and EPSRC through grant number EP/Y015673/1.
\end{acknowledgments}


\begin{appendix}

\section{The Unmodulated Case}
\label{sec:static}

The unmodulated scenario consists of studying a slab of material with thickness $L$ and permittivity $\varepsilon_0 = n_0^2$ embedded in vacuum ($\varepsilon^\pm=1$). We consider normal incidence on non-magnetic materials. Maxwell's equations in the time-harmonic regime (with the $\mathrm{e}^{-\mathrm{i} \omega t}$ time dependence suppressed) reduce to the one-dimensional wave equation
\begin{equation}
\frac{\partial^2 E}{\partial x^2} + k^2 \varepsilon E = 0,
\end{equation}
where $k=\omega/c$.\\
The electric field can be written as 
\begin{equation}
E = 
\begin{cases}
r\mathrm{e}^{-\mathrm{i} k x} & \text{if } x<0, \\
a^+\mathrm{e}^{\mathrm{i} k n_0 x} + a^-\mathrm{e}^{-\mathrm{i} k n_0 x} & \text{if } 0\leq x \leq L, \\
t\mathrm{e}^{\mathrm{i} k x} & \text{if } x>L.
\end{cases}
\end{equation}
Applying the boundary conditions at $x =0$ and $x=L$, we obtain the linear system $M(\omega) \ket{\Phi} = 0$ with $\ket{\Phi} = (a^+, a^-)^T$ and
\begin{equation}
M(\omega) = 
\begin{pmatrix}
1+n_0 & 1-n_0 \\
(1-n_0) \mathrm{e}^{\mathrm{i} \omega n_0 L/c} & (1+n_0) \mathrm{e}^{-\mathrm{i} \omega n_0 L/c} 
\end{pmatrix}.
\end{equation}
Finding the eigenvalues $\omega_m$ of the system requires solving $\det M = 0$, which yields \cite{muljarov2011}
\begin{equation}
\omega_m = \frac{c}{n_0 L} \left(m\pi - \mathrm{i} \log\left(\frac{n_0+1}{n_0-1}\right)\right), \quad m \in \mathbb{Z}.
\label{eq:evs_slab_static}
\end{equation}
The right eigenvectors satisfying $M(\omega_m)\ket{\Phi_m} = 0$ are given by
\begin{equation}
\ket{\Phi_m} = 
\begin{pmatrix}
1 \\
\frac{n_0+1}{n_0-1}
\end{pmatrix}.
\end{equation}
The left eigenvectors satisfy $\bra{\Psi_m} M(\omega_m) = 0$ and are normalized such that the biorthogonality condition $\bra{\Psi_m}M'(\omega_m) \ket{\Phi_m} = 1$ is enforced, where the prime denotes differentiation with respect to $\omega$:
\begin{equation}
\bra{\Psi_m} = \mathrm{i} \frac{c}{2 n_0 (n_0+1) L}
\begin{pmatrix}
1 & (-1)^m
\end{pmatrix}.
\end{equation}
Applying the Keldysh theorem (\ref{eq:keldysh1}), we can explicitly calculate $R$ since both $M^{-1}$ and the series in are known in closed form. After some calculation, we find that it is a constant matrix given by
\begin{equation}
R = \frac{1}{2}
\begin{pmatrix}
\frac{1}{1+n_0} & 0 \\
\frac{1}{1-n_0} & 0
\end{pmatrix}.
\end{equation}

\section{Empty Temporal Lattice Approximation}
\label{sec:empty}

In the limit where the modulation vanishes, the eigenpairs in \eqref{eq:mat_eig} are $k_p = n_0 (\omega-\Omega p)/c$ and $e_{pq} = \delta_{pq}$, where $n_0=\sqrt{\varepsilon_0}$. Substituting these into \eqref{eq:slab_evp}, we obtain a block-diagonal matrix $\widetilde{S}$ with eigenvalues corresponding to the unmodulated slab shifted by integer multiples of the modulation frequency:
\begin{equation}
\widetilde{\omega}_{p,q} = \widetilde{\omega}_q + p\Omega, \quad p \in \mathbb{Z},
\end{equation}
with associated right and left eigenmodes $\ket{\widetilde{\phi}_{p,q}}$ and $\bra{\widetilde{\psi}_{p,q}}$. The static slab eigenvalues $\widetilde{\omega}_q$ are given by
\begin{equation}
\widetilde{\omega}_q = \frac{c}{n_0 L} \left[q\pi - \frac{\mathrm{i}}{2} \log\left(\frac{(n_0+n^+)(n_0+n^-)}{(n_0-n^+)(n_0-n^-)}\right)\right].
\label{eq:evs_slab_static1}
\end{equation}

\section{First-Order Perturbative Approximation}
\label{sec:pert}

Writing $S = \widetilde{S} + \Delta S$ and assuming the eigenvalues $\widetilde{\omega}_{p,q}$ are distinct, first-order perturbation theory \citep{cohen1991quantum} yields
\begin{equation}
{\omega}_{p,q} \simeq \widetilde{\omega}_{p,q} - \frac{\bra{\widetilde{\psi}_{p,q}} \Delta S(\widetilde{\omega}_{p,q}) \ket{\widetilde{\phi}_{p,q}}}{\bra{\widetilde{\psi}_{p,q}} \frac{\partial S}{\partial \omega}(\widetilde{\omega}_{p,q})\ket{\widetilde{\phi}_{p,q}}}.
\end{equation}

\section{Finite element formulation}
\label{app:fem}

We seek solutions of the form $E(x, t) = \sum_{q} u_q(x) e^{-i(\omega - q\Omega)t}$, leading to a coupled system of Helmholtz equations for the harmonic amplitudes $u_q(x)$, and consider a finite interval domain $\Theta$. The permittivity is decomposed in Fourier series as $\varepsilon(x, t) = \sum_{q} \varepsilon_q(x) e^{-i(\omega - q\Omega)t}$. The weak formulation is: find $\{u_q\} \in H^1(\Theta)$ such that for all test functions $v_q \in H^1(\Theta)$,
\begin{align}
\sum_{p,q}  (\omega - p\Omega)^2 \int_\Theta &\left[ \varepsilon_{p-q}(x)u_p(x) \, \overline{v_q(x)} \right. \notag \\
&\left. + \, \delta_{p,q} \nabla u_p(x) \cdot \nabla \overline{v_q(x)} \right] \, dx = 0
\end{align}
This yields a quadratic eigenvalue problem in $\omega$, which we discretize using continuous second order Lagrange finite elements with FEniCSx~\cite{dolfinx2024} and solve using the SLEPc eigenvalue solver~\cite{slepc2005}.\\
To simulate open boundaries, we implement perfectly matched layers (PMLs) using a complex coordinate stretching of the form $x \mapsto x + i\sigma(x)$, where $\sigma(x)$ is a smooth function supported in the top and bottom PML region. This approach effectively introduces complex-valued coefficients in the weak form, allowing for the absorption of outgoing waves without reflections~\cite{chew1994pml}.

\end{appendix}
\bibliographystyle{apsrev4-2}
\bibliography{biblio}

\end{document}